\documentclass[conference]{IEEEtran}

\usepackage[english]{babel}
\usepackage{cite}
\usepackage{xcolor} 
\definecolor{lfdblack}{HTML}{000000}
\definecolor{lfdyellow}{HTML}{E69F00}
\definecolor{lfddgrey}{HTML}{999999}
\definecolor{lfdgreen}{HTML}{009371}
\definecolor{lfdhgrey}{HTML}{beaed4}
\definecolor{lfdred}{HTML}{ed665a}
\definecolor{lfdblue}{HTML}{1f78b4}
\definecolor{bggray}{gray}{0.9}
\usepackage{booktabs}
\usepackage{csquotes}
\usepackage{tablefootnote}
\usepackage{numprint}
\usepackage{nicematrix}
\usepackage[compat=0.6]{yquant}
\usepackage{lipsum}
\usepackage{comment}
\usepackage{subcaption}
\usepackage{pgfplots}
\pgfplotsset{compat = newest}
\usepackage{tikz}
\usetikzlibrary{arrows.meta, shapes.geometric,
                positioning,fit,shadows.blur,
                patterns,calc, matrix,
                tikzmark, backgrounds}
\usepackage{fp}
\usepackage{xspace}
\usepackage{amsmath,amssymb,amsfonts}
\usepackage{amsthm} \usepackage{bbm}
\usepackage{quantikz}
\usepackage{placeins}
\usepackage[T1]{fontenc}
\usepackage[colorlinks=true,urlcolor=teal,
            linkcolor=teal,citecolor=teal]{hyperref}
\usepackage[most]{tcolorbox}
\usepackage{listings}
\usepackage{censor}
\usepackage{todonotes}
\usepackage[printonlyused, nolist]{acronym}
\usepackage[bold,full]{complexity}

\newboolean{anonymous}
\setboolean{anonymous}{false}

\addto\extrasenglish{  %
}

\newcommand{\reprolink}{https://github.com/lfd/RL_for_JO}
\newcommand{\ie}{\emph{i.e.}\xspace}
\newcommand{\eg}{\emph{e.g.}\xspace}
\newcommand{\etal}{\emph{et al.}\xspace}

\makeatletter
\def\calcLength(#1,#2)#3{%
  \pgfpointdiff{\pgfpointanchor{#1}{center}}%
  {\pgfpointanchor{#2}{center}}%
  \pgf@xa=\pgf@x%
  \pgf@ya=\pgf@y%
  \FPeval\@temp@a{\pgfmath@tonumber{\pgf@xa}}%
  \FPeval\@temp@b{\pgfmath@tonumber{\pgf@ya}}%
  \FPeval\@temp@sum{(\@temp@a*\@temp@a+\@temp@b*\@temp@b)}%
  \FProot{\FPMathLen}{\@temp@sum}{2}%
  \FPround\FPMathLen\FPMathLen5\relax
  \global\expandafter\edef\csname #3\endcsname{\FPMathLen}
}
\makeatother

\lstdefinestyle{query}{
  language=SQL,
    stepnumber=1,
  numbersep=10pt,
  tabsize=4,
  showspaces=false,
  showstringspaces=false,
  basicstyle=\linespread{1}\fontfamily{lmtt}\selectfont\small,
  keywordstyle=\color{blue},
  stringstyle=\color{purple},
  upquote=true,
  breaklines=true,
  commentstyle=\color{CadetBlue}
}

\definecolor{mygray}{rgb}{0.643,0.643,0.643}
\newtcolorbox{querybox}[2][]{%
  sidebyside align=top,
  enhanced,
  boxsep=2pt,
  arc=0pt,
  top=-3pt, bottom=-3pt,
  left=2pt, right=0pt,
  colback=white,
  colframe=mygray,
  boxrule=0.5pt,
  leftrule=12pt,
  overlay unbroken and first ={%
      \node[rotate=90,
        minimum width=0.5cm,
        anchor=south,
                yshift=-11pt,
        white]
      at (frame.west) {#2};
    }
}

\newtcolorbox{matrixbox}[2][]{%
  sidebyside align=top,
  enhanced,
  boxsep=0pt,
  arc=0pt,
  left=-1em,
  top=-0.8em,
  boxrule=0pt,
  colframe=bggray,
  colback=bggray,
  leftrule=12pt,
  overlay unbroken and first ={%
      \node[rotate=90,
        minimum width=0.5cm,
        anchor=south west,
        font=\itshape,
        yshift=0pt,
        xshift=0.5em,
        black]
      at (frame.south west) {#2};
    }
}
\newcommand{\CGate}[1]{\ensuremath{\text{C\raisebox{0.08em}{--}}\!#1}\xspace}
\newcommand{\rx}{R_X}

\newcommand{\rz}{R_Z}
\newcommand{\rzz}{R_{ZZ}}
\newcommand{\rzzT}{R_{Z^2}}
\newcommand{\rzn}{R_{Z^n}}

\newcommand{\suppweb}{\href{https://github.com/lfd/qsw24-reduction-methods}{supplementary website}\xspace}
\newcommand{\repropkg}{\href{https://doi.org/10.5281/zenodo.11402739}{reproduction package}\xspace}

\ifbool{anonymous}{}{\renewcommand{\censor}[1]{#1}}
\ifbool{anonymous}{}{\renewcommand{\blackout}[1]{#1}}
\ifbool{anonymous}{\newcommand{\genemail}[2]{\href{mailto:xxx.xxx@xx.xx}{\blackout{#2}}}}{\newcommand{\genemail}[2]{\href{#1}{#2}}}

\begin{document}
\bstctlcite{BSTcontrol}

\begin{acronym}[qaoa]
  \acro{qaoa}[QAOA]{Quantum Approximate Optimisation Algorithm}
  \acro{qubo}[QUBO]{Quadratic Unconstrained Binary Optimisation}
  \acro{pubo}[PUBO]{Polynomial Unconstrained Binary Optimisation}
  \acro{wlog}[WLOG]{Without loss of generality}
  \acro{nisq}[NISQ]{Noisy Intermediate Scale Quantum}
  \acro{tsp}[TSP]{Travelling Saleseman Problem}
  \acro{pbf}[PBF]{Pseudo-Boolean Function}
  \acro{op}[OP]{Optimisation Problem}
  \acro{rmd}[RMD]{\emph{reduce/map/decompose}}
  \acro{md}[MD]{\emph{map/decompose}}
  \acro{nisq}[NISQ]{Noisy Intermediate Scale Quantum}
  \acro{vqe}[VQE]{Variational Quantum Eigensolver}
\end{acronym}

\title{Polynomial Reduction Methods\\ and their Impact on QAOA Circuits}
\author{
  \IEEEauthorblockN{\blackout{Lukas Schmidbauer}}
  \IEEEauthorblockA{\blackout{\textit{Technical University of}}\\
    \blackout{\textit{Applied Sciences Regensburg}} \\
    \blackout{Regensburg, Germany} \\
    \genemail{mailto:lukas.schmidbauer@othr.de}{lukas.schmidbauer@othr.de}}
  \and
  \IEEEauthorblockN{\blackout{Karen Wintersperger}}
  \IEEEauthorblockA{
    \blackout{\textit{Siemens AG, Technology}}\\
    \blackout{Munich, Germany}\\
    \genemail{mailto:karen.wintersperger@siemens.com}{karen.wintersperger@siemens.com}}
  \and
  \IEEEauthorblockN{\blackout{Elisabeth Lobe}}
  \IEEEauthorblockA{
    \blackout{\textit{German Aerospace Center}}\\
    \blackout{\textit{(DLR), Institute of Software}}\\
    \blackout{\textit{Technology, Department}}\\
    \blackout{\textit{High-Performance Computing}}\\
    \blackout{Braunschweig, Germany}\\
    \genemail{mailto:elisabeth.lobe@dlr.de}{elisabeth.lobe@dlr.de}}
  \and
  \IEEEauthorblockN{\blackout{Wolfgang Mauerer}}
  \IEEEauthorblockA{\blackout{\textit{Technical University of}}\\
    \blackout{\textit{Applied Sciences Regensburg}}\\
    \blackout{\textit{Siemens AG, Technology}}\\
    \blackout{Regensburg/Munich, Germany}\\
    \genemail{mailto:wolfgang.mauerer@othr.de}{wolfgang.mauerer@othr.de}}
}

\maketitle

\begin{abstract}
  Abstraction layers are of paramount importance in software architecture,
  as they shield the higher-level formulation of payload computations from lower-level details.
  Since quantum computing (QC) introduces many such details that
  are often unaccustomed to computer scientists, an obvious desideratum is to devise appropriate abstraction
  layers for QC. For discrete optimisation, one such abstraction is to
  cast problems in quadratic unconstrained binary optimisation (QUBO) form,
  which is amenable to a variety of quantum approaches. However, different mathematically equivalent forms can lead to different behaviour on quantum hardware, ranging from ease of mapping onto
  qubits to performance scalability.

  In this work, we show how using higher-order problem formulations
  (that provide better expressivity in modelling optimisation tasks
  than plain QUBO formulations) and their automatic transformation
  into QUBO form can be used to leverage such differences to prioritise
  between different desired non-functional properties for quantum
  optimisation.
  Based on a practically relevant use-case and a graph-theoretic
  analysis, we evaluate how different transformation approaches influence widely used
  quantum performance metrics (circuit depth, gates count,
  gate distribution, qubit scaling), and also consider the
  classical computational efforts required to perform the
  transformations, as they influence possibilities for achieving
  future quantum advantage. Furthermore, we establish more general properties
  and invariants of the transformation methods.
  Our quantitative study shows that the approach allows
  us to satisfy different trade-offs, and suggests various
  possibilities for the future construction of general-purpose
  abstractions and automatic generation of useful quantum circuits
  from high-level problem descriptions.
\end{abstract}


\section{Introduction}
\label{sec:intro}

Optimisation problems occur in the realm of practically relevant problems, such as finding optimal time-schedules, production planning, quality control, portfolio optimisation, social network analysis, protein folding and drug discovery.
Quantum computers promise complexity theoretical advantages in this
field, from quadratic acceleration for finding optimal elements in unstructured search spaces~\cite{Grover_1996} via possible speedups using adiabatic quantum
computing and annealing~\cite{Lidar:2018}, to in-principal super-polynomial speedups for approximate
combinatorial optimisation established using advanced reasoning~\cite{Pirnay:2024}.

\begin{figure}[htbp]
  \centering
  \includegraphics{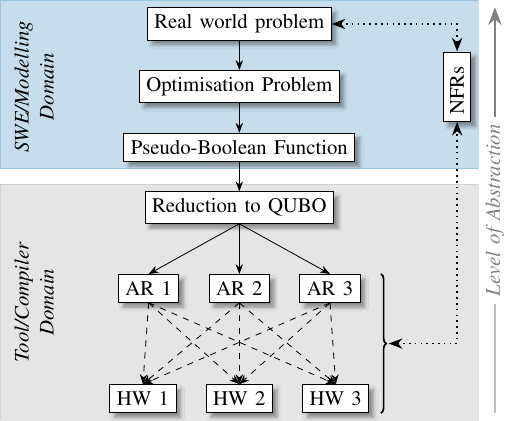}
  \caption{Influence of reduction process and quantum hardware properties
  on non-functional requirements (NFRs) across domains. AR denotes algebraic representations of a QUBO problem, HW represents transpilation results to different quantum backends. Dashed line: transpilation process, dotted line: dependency.}
\label{fig:SWHWAbstraction}
\end{figure}

Formulating an \ac{op} is an essential prerequisite to enjoying possible
advantages of large classes of quantum algorithms. On the one hand, one could
argue that such formulations are purely mathematical, and are therefore
a welcome abstraction layer that decouples the use of quantum algorithms
from knowledge of the inner working of quantum systems. On the other
hand, the performance of quantum algorithms usually strongly depends on the concrete formulation~\cite{Schoenberger_2023, Schoenberger_23:qdsm, schoenberger_2023_leapCoDesign, krueger_2020_SAT}.
In contrast to noiseless classical computers, this dependence is partly a consequence of current \ac{nisq} devices, which suffer from errors induced during execution---invalidating the quantum state~\cite{greiwe_23_imperfections, franz_2022_instabilities}.
Therefore, the question arises to what extent the degrees of freedom in the formulation of optimisation problems impact the performance of the quantum algorithms and thus to what extent these processes can be decoupled in the overlying design automation process.
\autoref{fig:SWHWAbstraction} illustrates the higher-level process of formulating a real world problem and transforming it to a suitable representation for quantum computers.
Here, dashed arrows abstract concrete procedures or algorithms, such as the \ac{qaoa}.
It also illustrates the present, yet undesirable, coupling of domains through non-functional requirements.

\ac{qaoa} is a widely used approach to solve combinatorial \ac{op}s \cite{Farhi_2014} on gate-based quantum computers.
To do so, the \ac{op}s are transformed into \ac{qubo} problems, or equivalently, except for an affine transformation, Ising models~\cite{Bian_2010, McClean_2016}
(apart from \ac{qaoa}, it is well known that other
approaches like \ac{vqe}~\cite{Tilly_2022} or quantum
annealing (inspired) approaches~\cite{Hauke2020, Schoenberger_2023, sax_2020_approximate} can solve \ac{qubo} problems).
The goal of a \ac{qubo} problem is to find $\vec{x}^* \in \{0,1\}^n$ which solves
\begin{equation}
  \begin{aligned}
    \text{min } & f(\vec{x}) \text{ such that } \vec{x} \in \{0, 1\}^n
  \end{aligned}
\end{equation}
for a given \ac{pbf} $f : \{0, 1\}^n \to \mathbb{R}$, where $f$ has a degree of at most 2.
The generalisation for functions $f$ of arbitrary degree is called \ac{pubo}.

\begin{figure}[htb]
  \centering
  \includegraphics{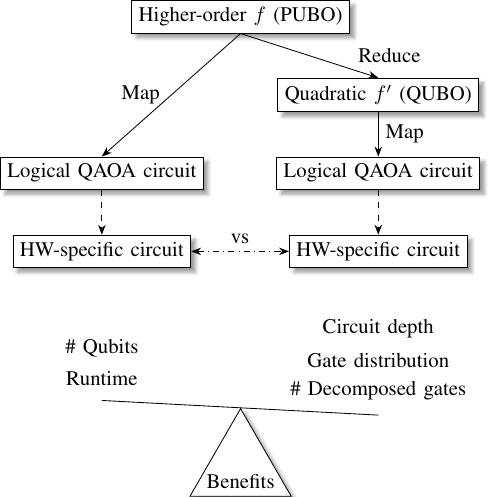}
  \caption{Overview of reduction strategies for higher-order unconstrained problems $f$
    to QAOA, and their consequences for execution on quantum hardware. Dashed line: Decomposition part of the transpilation process.}
  \label{fig:ResearchQuestionOverview}
\end{figure}

For practically relevant problem sizes, it is hard to stay within the restrictions imposed by noise and qubit count
of \ac{nisq}-devices.
Nonetheless, it is important to extrapolate findings beyond current HW capabilities to assess their scaling behaviour.
Thus, we use metrics, namely the circuit depth, the number of introduced gates, the gate distribution, the number of qubits and the runtime of reductions to analyse the performance of quantum algorithms.
Furthermore, current hardware natively supports one- and two-qubit gates. Higher-order gates can be executed via decompositions into a multitude of natively supported gates. Starting from an arbitrary \ac{pbf} $f$ (\ie, a \ac{pubo}) that encodes an \ac{op} analogously to above, we investigate two variants to reduce it to a QUBO and arrive at a \ac{qaoa} circuit, as illustrated by the flow diagram in \autoref{fig:ResearchQuestionOverview} (which can
also be seen as a more detailed view on the abstraction
layer in \autoref{fig:SWHWAbstraction}).

We focus on the problem Hamiltonian $H_P$ of a single layer, since $H_P$ depends on the encoded \ac{op} and is therefore the decisive part of our analysis (we deliberately ignore $H_M$, as it would only increase the
circuit depth by one).  The otherwise $p$-layer deep \ac{qaoa} circuit~\cite{Farhi_2014, Hadfield_2019} produces the state
\begin{equation}
  \label{eq_qaoa_parameterized_circuit}
  \ket{\vec{\beta}, \vec{\gamma}} = e^{-i\beta_p H_M} e^{-i \gamma_p H_P} \cdots\; e^{-i\beta_1 H_M} e^{-i \gamma_1 H_P} \ket{s}.
\end{equation}
The first variant, \ac{md} (\autoref{fig:ResearchQuestionOverview}: left path), directly encodes terms provided by $f$ into gates in the circuit and then starts a decomposition step.
This comes with the downside of higher circuit depth and more gates in the circuit.
The second variant, \ac{rmd} (\autoref{fig:ResearchQuestionOverview}: right path), order-reduces $f$ with
\(\operatorname{deg}(f) > 2\) to a quadratic \ac{pbf} $f'$ (\ie, a \ac{qubo}), maps terms from $f'$ to the circuit and then decomposes them to the hardware gate-set.
A reduction introduces additional variables, and therefore
increases the number of required qubits.
To avoid burdening the comparison with hardware-specific
details, we assume that the hardware gate-set consists of single-qubit $\rz$, $\rx$ operations, and the
two-qubit gate \CGate{X}.

This leads to our research questions:

\begin{description}
  \item[RQ1] How do circuits based on higher-order functions originating from \emph{map/decompose} (MD) and \emph{reduce/map/decompose} (RMD) differ
        in (a) depth, (b) size,  and (c) structure?
  \item[RQ2] What are the implications on the quantum software regarding abstraction from hardware specific peculiarities?
\end{description}

The rest of this paper is structured as follows: We first introduce the
fundamental  concept of \emph{quadratisation} in \autoref{sec:rel_work},
followed by details on formal representation and properties of
\ac{pbf}s in \autoref{sec:Fundamentals}, which also
introduces a graph representation to visualise \ac{pbf}s
and the effect of reductions.
The experimental analysis in \autoref{sec:Experiments} is
based on a practical use case, a scheduling problem, to
understand important performance metrics of quantum circuits
for strategies \ac{md} and \ac{rmd}.
We conclude by discussing the impact of our findings on
quantum software in \autoref{sec:effect_on_qc}.

The paper is augmented by a \suppweb and a comprehensive \repropkg~\cite{mauerer_22_QSaner} (links in PDF) that allows for extending our work.

\section{Related Work}\label{sec:rel_work}
Improving the performance of quantum algorithms is an intensively studied topic at many levels of abstraction.
Published findings concern issues from high-level software engineering to low-level design of compilers~\cite{Schmale_2022}.
Often, well established concepts from classical computing are evaluated in the quantum computing context.
Ahmad~\etal~\cite{Ahmad_2022} investigate modular architecture centric designs.
Scheerer~\etal~\cite{Scheerer_2022} evaluate architectural patterns for fault-tolerant systems. Faro~\etal~\cite{Faro_2023} and
Saurabh~\etal~\cite{Saurabh_2023} propose techniques to integrate quantum computation into data centres by the means of existing system architectures.
At a lower level, Codognet~\etal~\cite{Codognet_2022} compare annealing(-inspired) approaches, and
Schmale~\etal~\cite{Schmale_2022} review compiler phases for trapped ion devices.
Often, changes in one domain affect the performance in a lower domain.
Guimares and Tavares~\cite{Guimaraes_2022} Safi~\etal~\cite{Safi_2023} and Wintersperger~\etal~\cite{wintersperger:22:codes} address the HW-SW co-design aspects of quantum computing, which tries to balance restrictions in the realisability of hardware and positive effects in the software domain, such as qubit connectivity or error proneness of \ac{nisq} devices.

In principle it is possible to realise interaction gates involving more than two qubits, for example, with trapped-ion quantum computers~\cite{Monz_2009, Shapira_2020}. However, this is currently not possible on other platforms such as superconducting qubits, and not provided by any commercial quantum
devices.
Nonetheless, higher order gates can be replaced by decomposing them into multiple smaller gates that lead to the same quantum operation.
For \ac{qaoa}, an interesting decomposition is concerned with rotation gates around the $Z$-axis.
Suppose the single-qubit $\rz$ gate can be implemented natively.
Then the two-qubit $\rzz \vcentcolon = \rzzT$ gate can be decomposed into a symmetric arrangement of two \CGate{X}-gates around a single-qubit $\rz$ gate.
This method can be extrapolated to $\rzn$ gates and is motivated by the works of Campbell and Dahl~\cite{Campbell_2021}, with reference to
ZX-calculus~\cite{Cowtan_2014}.
They analyse the effect of decomposition for the four corner graph colouring problem.
They were able to execute small instances of \ac{qaoa} with COBYLA---a classical optimiser for $\vec{\beta}$ and $\vec{\gamma}$ (see \autoref{eq_qaoa_parameterized_circuit})---and found better performance for the mere decomposition strategy (\ac{md}), which differs from our findings.
They suggest that the use of higher order terms can be beneficial, which motivates our work.

In contrast to the \ac{md} strategy, one can also modify the \ac{op} so that higher-order gates are no longer needed. This is also interesting in view of using quantum annealing, as currently available devices can only handle interactions between a maximum of two qubits~\cite{Hauke2020}.

One particular approach is \emph{quadratisation}~\cite{Boros_2019}, which eliminates the need for $\rzn, n > 2$ gates.
A function $f'(\vec{x},\vec{y})$ is a \emph{quadratisation} of $f(\vec{x})$, if $f'(\vec{x},\vec{y})$ is a quadratic ($\operatorname{deg}(f') = 2$) \ac{pbf} in $\vec{x} = x_1,\ldots ,x_n$ and $\vec{y} = y_1,\ldots ,y_m$, and satisfies:
\begin{equation}
  \label{eq:QuadratizationMinimumPreserving}
  f(\vec{x}) = \min_{\vec{y}\in \{0,1\}^m} f'(\vec{x},\vec{y}) \; \forall \vec{x} \in \{0,1\}^n.
\end{equation}
Note that with this also the minimum over $\vec{x}$ is preserved.

A variety of methods, reviewed by Dattani~\cite{Dattani_2019}, are known to construct a quadratisation.
Typical approaches include rewriting $f(\vec{x}$) through logical reasoning of minima or exploiting substructures in $f(\vec{x})$ (\eg, via \emph{a priori} knowledge~\cite{Tanburn_2015}; splitting the objective function~\cite{Okada_2015}; excluding monomials~\cite{Tanburn_2015}; partial monomial assignment~\cite{Hiroshi_2014} or using Gröbner basis techniques~\cite{Dridi_2017}).
It is not uncommon that some methods require \(f\) to satisfy
particular properties (\eg, restricted degree~\cite{Gallagher_2011}; exclusively negative or positive monomials~\cite{Anthony_2016, Boros_2014} or non-binary variables~\cite{Rocchetto_2016}). A versatile method proposed by Boros~\cite{Boros_2002} quadratises an \emph{arbitrary} \ac{pbf} $f: \{0,1\}^n \rightarrow \mathbb{R}$ constructively by iteratively replacing a pair of variables $x_ix_j$ in the multi-linear representation of $f$ by a new variable $y_h$.
To  ensure the constraint specified
in~\autoref{eq:QuadratizationMinimumPreserving} and enforce $y_h = x_ix_j$ in the minimisation of $f$, a penalty $\operatorname{p}$ is added to the reduced \ac{pbf}, which fulfils
\begin{equation}
  \begin{split}
     & x_ix_j = y_h \Rightarrow \operatorname{p} = 0     \\
     & x_ix_j \neq y_h \Rightarrow \operatorname{p} > 0,
  \end{split}
\end{equation}
where \(\operatorname{p}\) is usually chosen as
\begin{equation}
  \operatorname{p}(x_i,x_j,y_h) = 3y_h + x_ix_j - 2x_iy_h -2x_jy_h.
\end{equation}

\section{Fundamentals}\label{sec:Fundamentals}

\subsection{Pseudo Boolean Functions and their Graph Representation}
\label{sec:PBF_and_graph_representation}
\label{sec:PropertiesPBF}
\paragraph{Pseudo Boolean Functions}

Following Boros~\etal~\cite{Boros_2019}, a \ac{pbf} \(f : \{0,1\}^n \rightarrow \mathbb{R}\) can be expressed in an
algebraic representation by multi-linear polynomials~\cite{Boros_2002}
\begin{equation}
  \label{eq:multi_linear_polynomial}
  f(x_1, \ldots , x_n) = \sum_{S\subseteq \{1,\ldots ,n\}} \alpha_S \prod_{j\in S}x_j,
\end{equation}
where $\alpha_S \prod_{j\in S}x_j$ is called a monomial of $f$ and $\alpha_S \in \mathbb{R}$.
For example, $f(x_1, x_2, x_3) = \pi x_1 + 3 x_1x_2 - 17x_2x_3$ is a \ac{pbf} and $\pi x_1$, $3 x_1x_2$ and $- 17x_2x_3$ are monomials of $f$.

In the following, we introduce terminology related to \ac{pbf}s.
Let $m_S$ be a monomial of the \ac{pbf} $f$, where $S \subseteq \{1,\ldots ,n\}$ is a subset of indices.
Then, we define the degree of $f$ by the maximum degree over $f$'s present ($\alpha_S \neq 0$) monomials:
\begin{equation}
  \operatorname{deg}(f) = \max_{S \subseteq \{1,\ldots ,n\}, \alpha_S \neq 0} \operatorname{deg}(m_S),
\end{equation}
where we define the degree of a monomial $m_S$ by the number of variables it contains:
\begin{equation}
  \operatorname{deg}(m_S) = |S| \in \{0,\ldots ,n\}.
\end{equation}
Note that $x^k = x$ for all $k \in \mathbb{N}$ for all binary variables $x \in \{0,1\}$.
Furthermore, we define the degree-$k$ ($k \in \mathbb{N}_0$) density of $f$ by the ratio $\frac{\text{actual}}{\text{possible}}$ degree-$k$ monomials.
Since there are $\binom{n}{k}$ possible degree-$k$ monomials\footnote{For the construction of a degree-$2$ monomial, there are $n(n-1)$ variable combinations. For a degree-$3$ monomial, there are $n(n-1)(n-2)$ combinations, etc.
  Since multiplication is commutative in each monomial, permutations, of which there are $k!$ many, are irrelevant.
  Consequently, there are $\frac{n!}{(n-k)!k!} = \binom{n}{k}$ ways to construct a degree-$k$ monomial.},
\begin{equation}
  \label{eq:DefinitionDensity}
  d_k = \frac{t_k}{\binom{n}{k}} \in [0,1],
\end{equation}
where $t_k$ denotes the number of actual terms of degree-$k$.
For example, the function $f(x_1, x_2, x_3) = 3 + 1x_1x_2 - 2 x_2x_3 + 7 x_1x_2x_3$ has degree $3$, since $\max \{\operatorname{deg}(m_{\{\}}), \operatorname{deg}(m_{\{1,2\}}), \operatorname{deg}(m_{\{2,3\}}), \operatorname{deg}(m_{\{1,2,3\}})\} = \max \{|\{\}|, |\{1,2\}|, |\{2,3\}|, |\{1,2,3\}|\} = 3$.
Furthermore, the degree-$k$ densities of $f$ are as follows:
\begin{equation}
  \begin{split}
    d_0 = \frac{1}{\binom{3}{0}} = 1           & ,\; d_1 = \frac{0}{\binom{3}{1}} = 0,                \\
    d_2 = \frac{2}{\binom{3}{2}} = \frac{2}{3} & ,\; d_3 = \frac{1}{\binom{3}{3}} = 1,\; d_{k>3} = 0.
  \end{split}
\end{equation}

\paragraph{Graph Representation for Pseudo Boolean Functions}
An undirected graph $G(V,E)$ is, as usual, given by vertices $v_i \in V$ and its edges $e = \{v_i, v_j\} \in E$.
We also consider multigraphs that can have duplicate edges so that $E$ becomes a multiset.
With reference to \autoref{eq:multi_linear_polynomial}, the set $V=\{v_1, \ldots , v_n\}$ is isomorphic to the set of variables in~$f$.
While $x_i$, for $i\in\{1,\ldots ,n\}$, is a variable in $f$, $v_i$ represents that variable in $G$.
The multiset of edges $E$ is constructed over $f$'s monomials.
Let $S\subseteq \{1,\ldots ,n\}$ be a subset of indices and let
\begin{equation}\label{eq:ps}
  P_S = \{\{v_a, v_b\} |\; a \in S \land b \in S\ \land a \neq b \land \alpha_S \neq 0 \}
\end{equation}
be the two-combination set of $S$.
Then $P_S$ is the set of edges generated by iterating over possible two-combinations of variables in a monomial specified by $S$.
Then, $E$ is the multiset-sum over all $P_S$.
For example, the graph $G(V,E)$ of \begin{equation}\label{eq:examplepbf}
  f(x_1,\ldots ,x_6) = 3 x_1 - 2 \textcolor{lfdyellow}{x_2x_3} + 5\textcolor{lfdblack}{x_1x_2x_3x_6} - 2\textcolor{lfdblue}{x_1x_2x_4x_5}
\end{equation}
is defined by $V = \{v_1, \ldots v_6\}$ and
\begin{equation*}
  \begin{aligned}
    E & = \sum_{S\subseteq \{1,\ldots ,6\}} P_S                                                                               \\
      & = \textcolor{lfdyellow}{P_{\{2,3\}}} + \textcolor{lfdblack}{P_{\{1,2,3,6\}}} + \textcolor{lfdblue}{P_{\{1,2,4,5\}}}   \\
      & = \{\textcolor{lfdyellow}{\{v_2,v_3\}},                                                                               \\
      & \phantom{{}={} \{}\textcolor{lfdblack}{\{v_1,v_2\}, \{v_1,v_3\}, \{v_1,v_6\}, \{v_2,v_3\}, \{v_2,v_6\}, \{v_3,v_6\}}, \\
      & \phantom{{}={} \{}\textcolor{lfdblue}{\{v_1,v_2\}, \{v_1,v_4\},\{v_1,v_5\}, \{v_2,v_4\}, \{v_2,v_5\}, \{v_4,v_5\}}\}.
  \end{aligned}
\end{equation*}

\begin{figure}[htb]
    \centering
    \begin{tikzpicture}
        \def\Biegeradius{5}
        \def\NodeRadius{1.5}

        \begin{scope}[circle]
            \draw
            (0.0:\NodeRadius)   node (1){$v_1$}
            (60.0:\NodeRadius)  node (2){$v_2$}
            (120.0:\NodeRadius) node (3){$v_3$}
            (180.0:\NodeRadius) node (4){$v_4$}
            (240.0:\NodeRadius) node (5){$v_5$}
            (300.0:\NodeRadius) node (6){$v_6$};
        \end{scope}

        \begin{scope}[-]
            \draw [lfdblue, densely dashed] (1) to [bend right = \Biegeradius] (2);
            \draw [lfdblack] (1) to [bend left  = \Biegeradius] (2);
            \draw [lfdblack] (1) to (3);
            \draw [lfdblack] (1) to (6);
            \draw [lfdblue, densely dashed] (1) to (4);
            \draw [lfdblue, densely dashed] (1) to (5);
            \draw [lfdyellow, dashdotted] (2) to [bend right = \Biegeradius] (3);
            \draw [lfdblack] (2) to [bend left  = \Biegeradius] (3);
            \draw [lfdblack] (2) to (6);
            \draw [lfdblue, densely dashed] (2) to (4);
            \draw [lfdblue, densely dashed] (2) to (5);
            \draw [lfdblack] (3) to (6);
            \draw [lfdblue, densely dashed] (4) to (5);
        \end{scope}
    \end{tikzpicture}
    \caption{Multigraph of \autoref{eq:examplepbf} before the reduction.
    }
    \label{fig:MultiGraphConnectivity_VarA}
\end{figure}
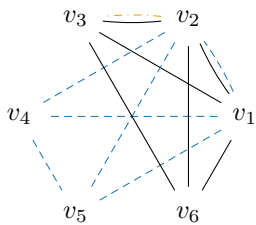

This graph can be seen in \autoref{fig:MultiGraphConnectivity_VarA}, where its vertices are drawn as the corners of a regular hexagon.
Here, multi-edges are drawn as separate edges in the graph, but can also be visualised by using an edge weight on a single edge.
Hence, an equivalent notation of (multi-)edges uses exponential notation: $\{v_i, v_j\}^\beta$ is called a multi-edge iff $\beta > 1, \beta \in \mathbb{N}$.

Notice that the graph representation is concerned with the structure of $f$, but not the numerical influence of a monomial of $f$, that is, the value of $\alpha_S$. Let $G(V,E)$ be the graph corresponding to a \ac{pbf} $f$.
If $G$ has multi-edges, then $\operatorname{deg}(f) > 2$ and equivalently, if $\operatorname{deg}(f) \leq 2$, $G$ has no multi-edges.
The inverse relation is not true in the general case.
An example is $f(x_1, x_2, x_3) = x_1x_2x_3$.
The corresponding graph $G$ is fully connected, but has no multi-edges, while $\operatorname{deg}(f) = 3$.
\autoref{sec:TheInfluenceOfReductions} discusses the implications for the reduction process.

\subsection{Variants of Boros Reduction}
Boros reduction~\cite{Boros_2002} can iteratively reduce a higher order \ac{pbf} $f$ to a quadratic one, while leaving the choice of the next variable pair during an iteration ambiguous (see \autoref{sec:rel_work}).
Although any variable pair in a monomial of $f$ is valid, a sensible choice involves characteristics of $f$.
We therefore evaluate the following three choices for the next variable:
\begin{enumerate}
  \item \textit{Sparse}: Gets the largest degree monomial. Chooses its first variable pair for the next iteration step
  \item \textit{Medium}: Gets all monomials with largest degree. Searches for the variable pair that appears most often.
  \item \textit{Dense}:  Searches for the variable pair that appears most among all monomials.
\end{enumerate}
The open source framework \href{https://gitlab.com/quantum-computing-software/quark}{\emph{quark}} (see Ref~\cite{Lobe_2023}) implements these methods. They
influence the reduction process and its outcome to the point where knowledge about the hardware topology can be used in advance to adjust the reduction process to optimise the hardware specific quantum circuit.
The variants are named according to the properties they
induce in the resulting \ac{qubo}.

\subsection{The Influence of Reductions}
\label{sec:TheInfluenceOfReductions}
\label{sec:influence_of_reduc}
Recall that Boros reduction introduces a penalty term in every iteration: $\operatorname{p}(x_i, x_j, y_h) = 3 y_h + x_ix_j - 2x_iy_h -2 x_jy_h$.

\paragraph{The Effect on multi-linear Polynomials}
Let $f: \{0,1\}^n \rightarrow \mathbb{R}$ be a \ac{pbf}, such that $\operatorname{deg}(f) > 2$.
A reduction results in a quadratic \ac{pbf} $f'$: $\operatorname{deg}(f') = 2$.
During the reduction, the new variables $y_1, \ldots , y_m$ are introduced, which adapt the domain of $f'$:
\begin{equation}
  \label{eq:EffectOnMultiLinearPolynomials_nm}
  f': \{0,1\}^{n+m} \rightarrow \mathbb{R}.
\end{equation}
Furthermore, the reduction process preserves the minimum of $f$ in $f'$, which follows from the quadratisation process of \autoref{eq:QuadratizationMinimumPreserving}.
This property lays the ground to the use of reductions in the reformulation of \ac{pbf}s.
Take into consideration that it might be necessary to scale the added term $\operatorname{p}$ with a large positive factor when solving $f'$ such that the penalty for breaking the constraint outweighs possible benefits in the objective function. This, however, does not influence the graph structure.

\paragraph{The Effect on the Graph Representation}
\begin{figure*}[htb]
    \centering
    \resizebox{\linewidth}{!}{%
        \begin{tikzpicture}
            \def\Biegeradiusa{3}
            \def\Biegeradiusb{10}
            \def\NodeRadius{1.6}
            \def\NodeRadiusHalf{0.8}

            \tikzstyle{rect1} = [rectangle,
            draw,
            align=center,
            minimum width=2cm,
            minimum height=.75cm,
            anchor=center];

            \newsavebox\GraphFirstn
            \sbox\GraphFirstn{\begin{tikzpicture}
                    \begin{scope}[circle]
                        \draw
                        (0.0:\NodeRadius)   node (x0){$v_0$}
                        (90.0:\NodeRadius)  node (x1){$v_1$}
                        (180.0:\NodeRadius) node (x2){$v_2$}
                        (270.0:\NodeRadius) node (x3){$v_3$};
                    \end{scope}

                    \begin{scope}[-]
                        \draw (x0) to [bend right = \Biegeradiusa] (x1);
                        \draw (x0) to [bend right = \Biegeradiusb] (x1);
                        \draw (x0) to [bend left = \Biegeradiusa] (x1);
                        \draw (x0) to [bend left = \Biegeradiusb] (x1);

                        \draw (x0) to [bend right = \Biegeradiusa] (x2);
                        \draw (x0) to [bend left = \Biegeradiusa] (x2);

                        \draw (x0) to [bend right = \Biegeradiusa] (x3);
                        \draw (x0) to [bend left = \Biegeradiusa] (x3);

                        \draw (x1) to [bend right = \Biegeradiusa] (x2);
                        \draw (x1) to [bend left = \Biegeradiusa] (x2);

                        \draw (x1) to [bend right = \Biegeradiusa] (x3);
                        \draw (x1) to [bend left = \Biegeradiusa] (x3);

                        \draw (x2) to (x3);
                    \end{scope}
                \end{tikzpicture}}

            \newsavebox\GraphSecondn
            \sbox\GraphSecondn{\begin{tikzpicture}
                    \begin{scope}[circle]
                        \draw
                        (0.0:\NodeRadius)   node (x0){$v_0$}
                        (90.0:\NodeRadius)  node (x1){$v_1$}
                        (180.0:\NodeRadius) node (x2){$v_2$}
                        (270.0:\NodeRadius) node (x3){$v_3$}
                        (45.0:\NodeRadiusHalf) node (y1){$y_1$};
                    \end{scope}

                    \begin{scope}[-]
                        \draw [lfdred] (x0) to (x1);
                        \draw [lfdred] (x0) to (y1);
                        \draw [lfdred] (x1) to (y1);

                        \draw (y1) to [bend right = \Biegeradiusa] (x2);
                        \draw (y1) to [bend left = \Biegeradiusa] (x2);

                        \draw (y1) to [bend right = \Biegeradiusa] (x3);
                        \draw (y1) to [bend left = \Biegeradiusa] (x3);

                        \draw (x2) to (x3);

                    \end{scope}
                \end{tikzpicture}}

            \newsavebox\GraphThirdn
            \sbox\GraphThirdn{\begin{tikzpicture}
                    \begin{scope}[circle]
                        \draw
                        (0.0:\NodeRadius)   node (x0){$v_0$}
                        (90.0:\NodeRadius)  node (x1){$v_1$}
                        (180.0:\NodeRadius) node (x2){$v_2$}
                        (270.0:\NodeRadius) node (x3){$v_3$}
                        (45.0:\NodeRadiusHalf)  node (y1){$y_1$}
                        (225:0.5) node (y2){$y_2$};
                    \end{scope}

                    \begin{scope}[-]
                        \draw (x0) to (x1);
                        \draw (x0) to (y1);
                        \draw (x1) to (y1);

                        \draw [lfdred] (y1) to (y2);
                        \draw [lfdred] (x2) to (y2);
                        \draw [lfdred] (x2) to (y1);

                        \draw (x3) to (y1);
                        \draw (x3) to (y2);
                    \end{scope}
                \end{tikzpicture}}
            \matrix (full) [matrix of nodes, row sep=1cm,column sep=2cm, ampersand replacement=\&] {
                \node (G1) [rect1] {\usebox{\GraphFirstn}}; \&
                \node (G2) [rect1] {\usebox{\GraphSecondn}}; \&
                \node (G3) [rect1] {\usebox{\GraphThirdn}};                                                                                                                                                                                         \\
                \node (f1) [rect1, text width= 5cm] {$f_1(x_0,x_1,x_2,x_3) = \textcolor{lfdblue}{x_0x_1} + \textcolor{lfdblue}{x_0x_1}x_2 + \textcolor{lfdblue}{x_0x_1}x_3 + \textcolor{lfdblue}{x_0x_1}x_2x_3$}; \&
                \node (f2) [rect1, text width= 5cm] {$f_2(x_0,x_1,x_2,x_3, y_1) = 4\textcolor{lfdred}{y_1} + \textcolor{lfdred}{x_0x_1} -2 \textcolor{lfdred}{x_0y_1} - 2 \textcolor{lfdred}{x_1y_1} + \textcolor{lfdblue}{y_1x_2} + y_1x_3 + \textcolor{lfdblue}{y_1x_2}x_3$}; \&
                \node (f3) [rect1, text width= 5cm] {$f_3(x_0,x_1,x_2,x_3, y_1, y_2) = 4y_1 + 4\textcolor{lfdred}{y_2} + \textcolor{lfdred}{y_1x_2} - 2\textcolor{lfdred}{y_2y_1} -2 \textcolor{lfdred}{y_2x_2} + x_0x_1 - 2x_0y_1 -2 x_1y_1 + y_1x_3 + y_2x_3$}; \\
            };

            \begin{scope} [-Stealth,black, thick]
                \draw (f1.east) -- node [above] {$x_0x_1 = y_1$} (f2.west);
                \draw (f2.east) -- node [above] {$y_1x_2 = y_2$} (f3.west);

                \draw (f1.north) -- node [above, sloped] {$G_1$} (G1.south);
                \draw (f2.north) -- node [above, sloped] {$G_2$} (G2.south);
                \draw (f3.north) -- node [above, sloped] {$G_3$} (G3.south);
            \end{scope}
        \end{tikzpicture}
    }
    \caption{A complete multi reduction evolution. Blue symbols represent the variable pair to be replaced; orange lines and symbols represent edges and terms introduced via the penalty term in the current step. All variable-selection types (\ie, \emph{Sparse}, \emph{Medium}, \emph{Dense}) lead to identical results, assuming multiple valid choices are resolved by choosing the first pair for each type.}
    \label{fig:monomial_multiReductionEvolutionComplete}
\end{figure*}
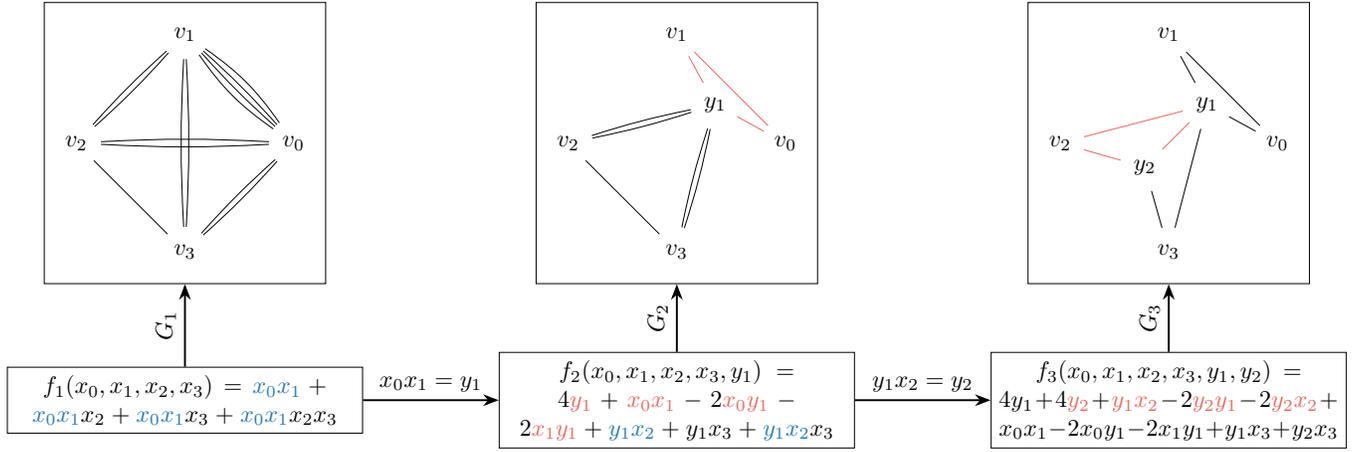
Let $f$ be a \ac{pbf}, such that its graph $G(V,E)$ contains multi-edges.
For example, \autoref{fig:monomial_multiReductionEvolutionComplete} shows such a function.
The multi-edge $\{v_0, v_1\}^4$ is replaced by a single edge (from the penalty term) in the first step.
Additionally, the penalty term connects the new node $y_1$ to the former nodes $v_0$ and $v_1$.
Due to $x_0x_1$ being replaced in every monomial by $y_1$, edges from these monomials are now mapped to $y_1$.

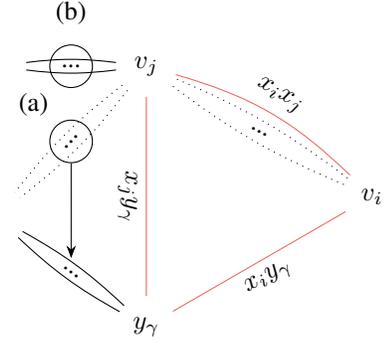
\begin{figure}[htb]
    \centering
    \begin{tikzpicture}
        \def\Biegeradius{7}
        \def\Biegeradiustwo{14}
        \begin{scope}[circle, minimum size = 0.8cm]
            \draw
            (0:2)   node[] (vi){$v_i$}
            (120:2) node[] (vj){$v_j$}
            (240:2) node[] (yn){$y_\gamma$};
            \draw node[draw, minimum size=0.5cm, left of = vj, label=above:{(b)}] (c3o) {...};
            \draw node[left of = c3o] (c3odummy) {};

            \draw node[draw, minimum size = 0.5cm, below of = c3o, rotate=45, label=90:{(a)}] (c4o){...};
            \draw node[left of = c4o] (c4odummy1) {};
            \draw node[below of = c4odummy1] (c4odummy) {};

        \end{scope}

        \begin{scope}[-]
            \draw [dotted] (vi) to [bend right = \Biegeradius]  node[below, sloped] {...} (vj);
            \draw [dotted] (vi) to [bend left  = \Biegeradius] (vj);

            \draw [lfdred] (vi) to [bend right = \Biegeradiustwo] node[lfdblack, above, sloped] {$x_ix_j$} (vj);
            \draw [lfdred] (vi) to node[lfdblack, below, sloped] {$x_iy_\gamma$} (yn);
            \draw [lfdred] (vj) to node[lfdblack, below, sloped] {$x_jy_\gamma$} (yn);

            %outer
            \draw (vj) to [bend right = \Biegeradius] (c3odummy);
            \draw (vj) to [bend left = \Biegeradius] (c3odummy);

            \draw [dotted] (vj) to [bend right = \Biegeradius] (c4odummy);
            \draw [dotted] (vj) to [bend left = \Biegeradius] (c4odummy);

            \draw [] (yn) to [bend right = \Biegeradius]  node[below, sloped, yshift=1] (c4target) {...} (c4odummy);
            \draw [] (yn) to [bend left = \Biegeradius] (c4odummy);

            \draw [-Stealth] (c4o) to (c4target);
        \end{scope}
    \end{tikzpicture}
    \caption{Effect of a reduction on $v_i$ and $v_j$. Dotted lines: Removed edges. Orange lines: Edges introduced by the penalty term. $\{v_i, v_j\}^\beta$ corresponds to $x_ix_j$, which is replaced by $y_\gamma$. The remapping of (a) ($x_ix_j \in m$) and the invariance of (b) ($x_ix_j \notin m$) is analogous for $v_i$ (not drawn).}
    \label{fig:ProofCategory1Visualization}
\end{figure}

More generally, suppose the variable selection type chooses the variable pair $x_ix_j$ and therefore the multi-edge $\{v_i,v_j\}^\beta$ in a \ac{pbf} $f_\gamma : \{0,1\}^n \rightarrow \mathbb{R}$ and its corresponding graph $G_\gamma(V_\gamma, E_\gamma)$ for iteration step $\gamma$.
The multi-edge $\{v_i,v_j\}^\beta$ is replaced by a single edge $\{v_i,v_j\}$, as every occurrence of $x_ix_j$ is replaced by $y_\gamma$ apart from the term introduced in the penalty term itself.
Additionally, the penalty term introduces $\{v_i, y_\gamma\}, \{v_j, y_\gamma\}$.
Let $m$ denote a monomial in $f_\gamma$.
Either (a) $x_ix_j \in m$ or (b) $x_ix_j \notin m$.
Let $I = \{\{v_\tau, v_i\} \in E_\gamma |\; \tau \in \{1,\ldots ,n\} \setminus \{j\}\}$ be the subset of edges connected to $v_i$ - excluding the ones connecting $v_i$ and $v_j$.
Analogously, let $J = \{\{v_\tau, v_j\} \in E_\gamma |\; \tau \in \{1,\ldots ,n\}\setminus \{i\}\}$ be the subset of edges connected to $v_j$ - excluding the ones connecting $v_i$ and $v_j$.
Every edge $e \in I \cup J$ stemming from (a) in $G_\gamma$ will be reconnected to $y_\gamma$ in $G_{\gamma+1}$.
Conversely, every edge $e \in I \cup J$ stemming from (b) in $G_\gamma$ is invariant in $G_{\gamma+1}$.
\autoref{fig:ProofCategory1Visualization} visualises the above stated, while the penalty term induced edges are coloured in orange.
The remaining edges, that is the ones not connected to $v_i$ or $v_j$ in $G_\gamma$, are invariant in $G_{\gamma +1}$.
This locality can be used for parallel execution of multiple reduction steps and thus aids in the speedup necessary for efficiently deploying the \ac{rmd} strategy.
Apart from parallel execution it lays the ground for an efficient data structure that does not need to reiterate over all monomials in search for the next variable pair in each reduction step.

\begin{figure}[htb]
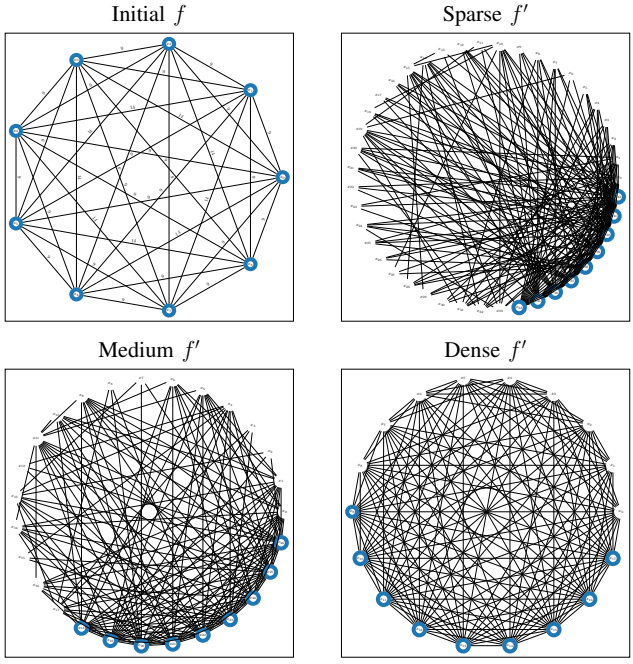

    \centering
    \resizebox{\linewidth}{!}{%
        \begin{tikzpicture}
            \def\Biegeradiusa{3}
            \def\Biegeradiusb{10}

            \def\InternalPlotSize{\linewidth}

            \tikzstyle{rect1} = [rectangle,
            draw,
            align=center,
            anchor=center];

            \tikzstyle{Annotation} = [
            font=\huge,
            yshift=-0.9cm
            ]

            \newsavebox\GraphBefore
            \sbox\GraphBefore{\resizebox{\InternalPlotSize}{\InternalPlotSize}{
                    \input{pics/GraphBefore.tex}}}

            \newsavebox\GraphAfterSparse
            \sbox\GraphAfterSparse{\resizebox{\InternalPlotSize}{\InternalPlotSize}{
                    \input{pics/3_3_35_Sparse.tex}}}

            \newsavebox\GraphAfterMedium
            \sbox\GraphAfterMedium{\resizebox{\InternalPlotSize}{\InternalPlotSize}{
                    \input{pics/3_3_19_Medium.tex}}}

            \newsavebox\GraphAfterDense
            \sbox\GraphAfterDense{\resizebox{\InternalPlotSize}{\InternalPlotSize}{
                    \input{pics/3_3_10_Dense.tex}}}

            \matrix (full) [matrix of nodes, row sep=1.5cm,column sep=1.5cm, ampersand replacement=\&] {
                \node (GB) [rect1] {\usebox{\GraphBefore}}; \& \node (GAS) [rect1] {\usebox{\GraphAfterSparse}};  \\
                \node (GAM) [rect1] {\usebox{\GraphAfterMedium}}; \& \node (GAD) [rect1] {\usebox{\GraphAfterDense}}; \\
            };

            \draw node[Annotation, above = of GB.north] (initf) {Initial $f$};
            \draw node[Annotation, above = of GAS.north] (initf) {Sparse $f'$};
            \draw node[Annotation, above = of GAM.north] (initf) {Medium $f'$};
            \draw node[Annotation, above = of GAD.north] (initf) {Dense $f'$};
        \end{tikzpicture}
    }
    \caption{A complete multi reduction evolution in terms of the graph structure. The symmetric graph in the top left corner shows the initial \ac{pbf} $f$ ($\operatorname{deg}(f) = 4$). The other nodes show a particular variant of Boros \cite{Boros_2002} reduction method starting from the initial \ac{pbf}. Nodes circled in blue represent the initial variables.}
    \label{fig:monomial_multiReductionEvolutionCompleteGraph}
\end{figure}

An edge in the graph corresponds to either a degree-$2$ monomial or a higher degree monomial.
Every edge, corresponding to a higher degree monomial is a potential candidate for the reduction.
The \textit{Dense} variable selection type restricts its choice to variable pairs that occur most often among all monomials - or equivalently the biggest multi-edges.
The \textit{Sparse} and \textit{Medium} variable selection types lift that constraint and thus might select other edges.
This is the only difference, yet it leads to vastly different properties of the resulting \ac{qubo}.
We observe differences in the edge distribution on nodes among the three variable selection variants concerning the graph representation.
While the \textit{Sparse} method concentrates its edges among the former variables, the \textit{Dense} method distributes its edges among all nodes, including newly introduced ones.
The \textit{Medium} method lies in between these two.
\autoref{fig:monomial_multiReductionEvolutionCompleteGraph} illustrates the graph representation for a \ac{pbf} $f$ ($deg(f) = 4$), in the top left corner and shows the graph for the resulting quadratic \ac{pbf} $f'$ for each variable-selection type.
The graph of a quadratic \ac{pbf} $f'$ is a representation of its \ac{qubo} (except for linear and constant terms), and thus describes coupling structure and density, since the pair-combination sets $P_S$ of $f'$ are pairwise disjoint.
Therefore, the properties of $f'$ in \autoref{fig:monomial_multiReductionEvolutionCompleteGraph} characterise the structure of the \ac{qubo}.
\autoref{sec:Experiments} discusses further differences that are relevant for our performance measurements.

Reductions leave some graph properties invariant:
\begin{enumerate}
  \item The total size of multi-edges in the graph strictly decreases.
  \item The degree of any node in the graph does not increase (excluding the new node in each iteration).
\end{enumerate}
While we defer formal proofs to
the \suppweb, note that they provide insights into the reductions:
Usually, the starting \ac{pbf} $f \vcentcolon = f_0$ ($\operatorname{deg}(f_0) > 2$) corresponds to a graph with multi-edges \footnote{\ac{pbf}s with pairwise disjoint sets $P_S$ do not induce multi-edges in their graph representation; for example, this is the case for $f(x_1,\ldots ,x_n) = x_1x_2x_3 + x_4x_5x_6 + \ldots + x_{n-2}x_{n-1}x_n$.}.
When using the \textit{Dense} variable-selection method, the algorithm at first reduces $f_0$ to $f_t$ after $t \in \mathbb{N}$ reduction iterations.
The graph corresponding to $f_t$ might not contain multi-edges anymore, however $f_t$ might still not be a quadratisation of $f_0$.
Hence, the algorithm enters stage two, where it can no longer select multi-edges and is therefore less efficient in terms of introduced variables, since it can no longer replace the same variable pair in multiple monomials.
We can calculate the amount of introduced variables in stage two for $f_t$ by evaluating each monomial.
A prerequisite of entering stage two is that the sets $P_S$ for $f_t$ are pairwise disjoint.
It is easy to see that for any monomial $m$ with $\operatorname{deg}(m) = k > 2$, we require $k - 2$ new variables to reduce it to degree 2.
Take for example the monomial $m = x_1x_2x_3$.
Here, one extra variable is needed for the quadratisation of $m$ regardless of the choice of the variable pair.

\section{Experiments}
\label{sec:Experiments}
\subsection{Experimental Setup}
Our experiments evaluate a \ac{pbf} $f$ that encodes higher-order terms and can scale in the number of variables.
It is motivated by an industrial job-shop scheduling problem: Production jobs must be assigned to machines such that the total runtime of a factory (the makespan) is minimised. For each job, only a single operation can be performed, and all machines are of the same type, so each job can in principle be run on each machine. The jobs require certain tool setups and are divided into multiple setup groups. Changing the setup takes additional time, which constrains optimal assignment. Here, we only aim for the allocation of jobs to machines, and leave the exact ordering jobs to classical post-processing. The problem is modelled as a \ac{pubo} using binary variables $x_{ij}$ with
\begin{equation}
  \label{eq:sag_var_part1}
  x_{ij} = \begin{cases}
    1 & \text{if job}\, i \,\text{is assigned to machine}\, j, \\
    0 & \, \text{otherwise.}
  \end{cases}
\end{equation}
This yields a problem size of $n = N \cdot M$ qubits for $N$ jobs and $M$ machines.
The goal is to minimise the maximum runtime including setup change times over all machines, which is expressed as minimising the differences between runtimes over all pairs of machines. Normally, there are many more jobs than machines and the job durations $d_i$ are longer than the setup change times. The setup change times between two jobs $i$ and $i'$ are given by the entries of the $N \times N$ matrix $R_{ii'}$, while another $N \times M$ matrix $S_{ij}$ describes the setup change time between a job $i$ and the initial setup of machine $j$.
Using the variables from Eq.~\ref{eq:sag_var_part1}, the objective term is given by
\begin{align}
  H_{\mathrm{obj}} & = \sum_{j < j'} \left[ \left(\sum_{i=1}^N x_{ij} (d_i+S_{ij}) + \sum_{i<i'} x_{ij} x_{i'j} R_{ii'} \right) \right. \\ \nonumber
                   & - \left. \left(\sum_{i=1}^N x_{ij'} (d_i+S_{ij'}) + \sum_{i<i'} x_{ij'} x_{i'j'} R_{ii'} \right) \right]^2.
  \label{eq:obj1}
\end{align}
As the order of jobs is not considered, we evaluate the maximum number of setup changes here. To avoid that the runtimes of two machines are equalised by adding additional, unwanted setup changes, we add a second term that minimises the setup changes on each machine separately:
\begin{align}
  H_{r} = \sum_{j=1}^M \left[\sum_{i < i'}^N x_{ij} x_{i'j} R_{ii'} + \sum_{i=1}^N x_{ij} S_{ij}\right].
\end{align}
Moreover, each job has to be assigned exactly once which is described by the following constraint:
\begin{align}
  H_{\mathrm{single}} = \sum_{i=1}^N \left( \sum_{j=1}^M x_{ij} -1 \right)^2.
  \label{eq:c1}
\end{align}
The \ac{pbf} $f: \{0,1\}^n \rightarrow \mathbb{R}$ ($\operatorname{deg}(f) = 4$) arises naturally from these preliminary steps as
\begin{equation}
  \label{eq:JSS_pbf_explicit}
  f = A H_\text{obj} + B H_r + C H_\text{single},
\end{equation}
where $C > A, B$ must hold to ensure valid solutions are preferred.
We use $A=1$, $B = 2\cdot\max_{i} d_i$, $C = 4 \cdot \max_{i}d_i^2$ and choose $R$ and $S$ to have full rank.

For each of  \autoref{fig:JointVarsAfterRuntime}, \ref{fig:Densities} and \ref{fig:JointCircDepthIntrGates},
the x axis shows the number of variables before the reduction of $f$, that is, problem size $n$.
Dependent on this characteristic, we evaluate run-time and variable overhead of the proposed reduction variants.
Additionally, we compare number of introduced gates, circuit depth and polynomial structure for both paths of \ac{md} and \ac{rmd}.
For this, we create the problem-specific part of the \ac{qaoa} circuit for a single layer without mixer.
To compare both paths, we decompose $\rzn$,  $n \geq 2$ gates into their symmetric representation given by Campbell and Dahl~\cite{Campbell_2021}.
Here, we do not optimise for circuit depth.
While this influences the comparison of circuit depths, we
note optimising the depth of a quantum circuit is in itself \NP{}-complete~\cite{Majumdar_2021}.

\subsection{Experimental Results}
\begin{figure}[b!]
    \centering
    \input{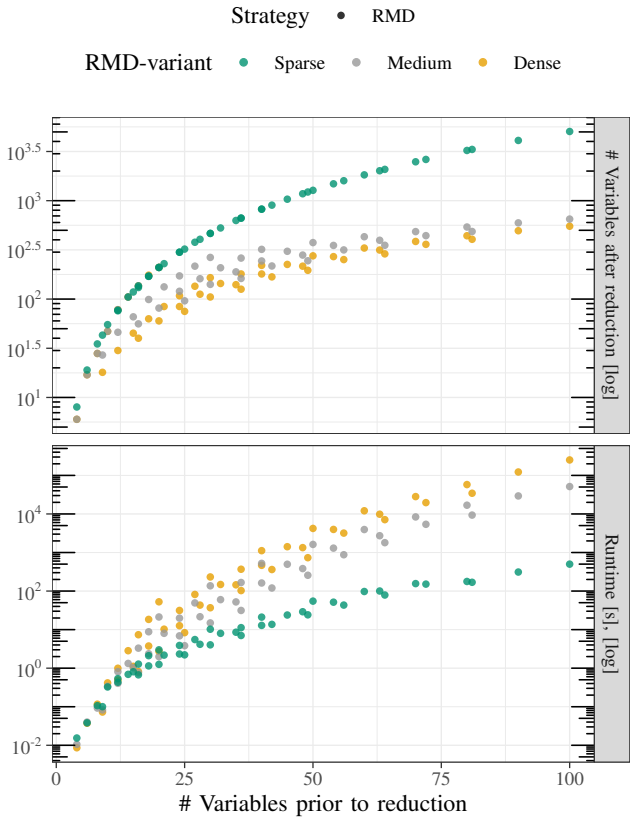}
    \caption{Top: Number of qubits (\ie variables) after the reduction over  problem size as captured by the number of variables before the reduction used to assess the spatial feasibility (\emph{reduce/map/decompose} (RMD) strategy).
    Bottom: Runtime (pure reduction) over problem size and reduction type for the \ac{rmd} strategy to assess computational feasibility.}
    \label{fig:JointVarsAfterRuntime}
\end{figure}%
\autoref{fig:JointVarsAfterRuntime} (bottom) shows the runtime for each variable-selection type of Boros \cite{Boros_2002} reduction method.
While the total number of steps for a reduction, that is the number of introduced variables, can be estimated using $f$'s multi-graph (see \autoref{sec:influence_of_reduc}), the total runtime also depends on the search and replacement implementation.
The latter one is shared among the variable-selection variants.
\autoref{fig:JointVarsAfterRuntime} (bottom) shows a diverging runtime trajectory between the \textit{Sparse} and \textit{Dense} variable-selection type.
This traces back to the search strategy as the dominant part for the next variable selection type.
The \textit{Sparse} variant introduces more variables, as \autoref{fig:JointVarsAfterRuntime} (top) illustrates, which means that the replacement occurs more often, which increases the runtime\footnote{The replacement occurs on different polynomials, which affects the runtime. We estimate this effect to be marginal for the given implementation.} on the one hand.
On the other hand, the \textit{Sparse} variant has a far less computationally intensive search strategy, which results in an overall significantly lower runtime, compared to the \textit{Dense} variant.
Since the \textit{Medium} selection-type is computationally less expensive than the \textit{Dense} type concerning the search strategy, we expect a lower runtime.
\autoref{fig:JointVarsAfterRuntime} confirms this.

Apart from the computational feasibility of the reductions, their spacial influence on \ac{pbf}s is relevant to quantum computing, as the number of variables translates to the number of qubits in the quantum circuit.
\autoref{fig:JointVarsAfterRuntime} (top) shows the dependency of the number of variables after the reduction (y-axis) on the the number of variables before the reduction (x-axis) grouped by the variable-selection type.
For the absolute values, the underlying polynomial and its structure is decisive, since it determines the size and coupling density of the corresponding multi-graph $G(V,E)$.
In contrast to the runtime, the variable-overhead mirrors the roles of the \textit{Sparse}, \textit{Medium} and \textit{Dense} types.
While the \textit{Sparse} type took the least amount of time, it introduces significantly more variables.
The reason behind this is the less efficient selection type, which chooses smaller (multi-)edges on average in the graph representation during the reduction process.
The upper bound for the number of introduced variables during a reduction is given by the polynomial's degree.
To be more precise, let $f: \{0,1\}^n \rightarrow \mathbb{R}$ be the function of interest, such that $\operatorname{deg}(f) = i$.
There are at most $\binom{n}{k} = \mathcal{O}(n^k)$ degree-$k$ monomials (see \autoref{sec:PropertiesPBF}).
At maximum, one needs $k-2$ extra variables per degree-$k$ monomial (see \autoref{sec:TheInfluenceOfReductions}) and thus $\mathcal{O}(\sum^{i}_{k=3}k n^k) = \mathcal{O}(n^i)$ extra variables for the quadratisation of $f$.
In our case, that is $\operatorname{deg}(f) = 4$, the number of introduced variables is upper bounded by $\mathcal{O}(n^4)$.
Note that \autoref{fig:JointVarsAfterRuntime} (top) shows the sum of the number of introduced variables and variables before the reduction on the y-axis (\ie, $n+m$; see \autoref{eq:EffectOnMultiLinearPolynomials_nm}).

\begin{figure}[tb]
    \input{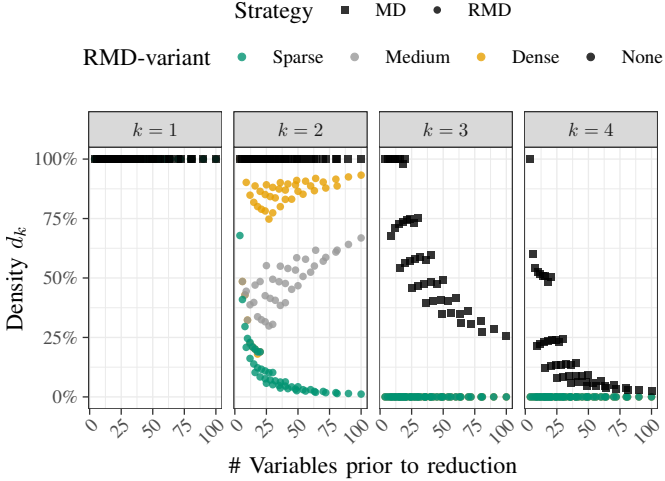}
    \caption{Polynomial density over problem size as measured by the number of variables prior to reduction. The graph shows the reduction from a higher order polynomial (\ac{md}) to a quadratic \ac{pbf} (\ac{rmd}) for different variable selection types.}
    \label{fig:Densities}
\end{figure}%
We now want to characterise the underlying function $f$ in terms of densities to get a better understanding of the above stated upper bound and to anticipate the influence on the quantum circuit.
\autoref{fig:Densities} shows the degree-$k$ densities $d_k$, as introduced in \autoref{sec:PropertiesPBF}, of $f$ before the reduction (\ac{md}) and of its quadratic counterpart $f'$ grouped by the variable-selection type (\ac{rmd}).
For the quadratic \ac{pbf} $f'$, $d_{k>2} = 0$, since there are no monomials $m$ in $f'$, such that $\operatorname{deg}(m) > 2$.
The density $d_1$ does not change in this case, since $f$ incorporates the maximum number of degree-$1$ terms already.
In each reduction step, a new variable is introduced.
At the same time, a unique degree-$1$ term is added to $f$ through the penalty term (see \autoref{sec:PropertiesPBF}).
For the general case, as the number of reduction steps increases, $d_1$ converges to the maximum value, that is $d_1 \to 1$, regardless of the number of degree-$1$ terms in the original function\footnote{The rationale being $\lim_{i\to\infty} \frac{t_1 + i}{\binom{n+i}{1}} = 1 \;\forall t_1 \in \{0,\ldots ,n\}$.}.
Interestingly, the convergence behaviour of $d_2$ is dependent on the variable-selection type.
\autoref{fig:Densities} shows this through our quantitative analysis.
While the \textit{Dense} selection type seems to converge to $d_2 > 90\%$, the \textit{Sparse} type converges to $d_2 < 10\%$.
Similar to \autoref{fig:JointVarsAfterRuntime}, the \textit{Medium} type lies in between the \textit{Sparse} and \textit{Dense} type.
This has immediate consequences for the quantum circuit, since there are more variables in case of the \textit{Sparse} method, but less two-qubit interactions between them, as indicated by the lower density. As not all possible pairs interact, the two-qubit gates can be parallelised more easily.
Take into consideration that the underlying function $f$ does not feature every degree-$4$ and degree-$3$ monomial.
In fact, the densities $d_3$ and $d_4$ converge to zero, which is caused by the choice of $f$.
We estimate, that our findings for the convergence hold for similar structured functions as well.
Recall the definition of $d_2$ from \autoref{eq:DefinitionDensity}.
Since
\begin{equation}
  \lim_{n\to\infty} d_2 = \lim_{n\to\infty} \frac{t_2}{\Theta(n^2)} =
  \begin{cases}
    0 & \text{if } t_2 \in o (n^2)      \\
    c & \text{if } t_2 \in \Theta (n^2)
  \end{cases},
\end{equation}
such that $0 < c \leq 1$,
we argue that, based on \autoref{fig:Densities}, the \textit{Dense} selection type is efficient in terms of encoding information.\footnote{This cannot be traced back to the penalty term, since it adds exactly $3$ edges per iteration (see \autoref{sec:influence_of_reduc}). It is therefore a linear function in terms of the iteration count and does not close the gap to the quadratic function.}

\begin{figure}[tb]
    \centering
    \input{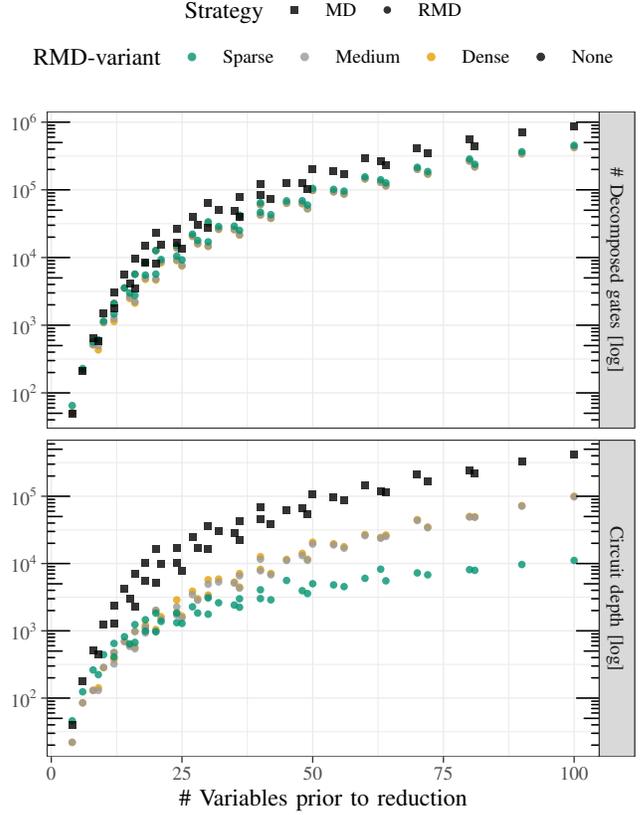}
    \caption{Single- and two-qubit gates (excluding
    gates that implement the mixer operation) in one 
    \ac{qaoa} layer over problem size as captured by the
    the number of variables for strategies \emph{map/decompose} (MD) and \emph{reduce/map/decompose} (RMD).
    Top: Gate count used as proxy for gate errors;
    bottom: Circuit depth (not optimised; gates inserted lexicographical) as proxy for decoherence effects.}
    \label{fig:JointCircDepthIntrGates}
\end{figure}%
The top part of \autoref{fig:JointCircDepthIntrGates} classifies the variable-selection types based on the number of introduced gates in a single \ac{qaoa} layer without the mixer.
The variable-selection types encode approximately the same information in their respective quadratic polynomials, since they introduce approximately the same number of gates.
In terms of the number of introduced gates in a \ac{qaoa} circuit, the \ac{rmd} strategy is superior to the \ac{md} strategy.
Take into consideration, the decomposition of higher-order gates influences this metric.
The number of gates is an important measure in the design of quantum circuits, as the cumulative effect of gate-errors invalidates the quantum state in the \ac{nisq}-era.

In contrast to the bare number of gates, the circuit depth takes into account possible parallel execution of gates. It determines the quantum algorithm's runtime and therefore the extent of decoherence effects. The bottom part of
\autoref{fig:JointCircDepthIntrGates} visualises circuit depth, where the \ac{rmd} strategy features a lower circuit depth than the \ac{md} strategy.
In short, the \ac{rmd} strategy suffers from more variables, while at the same time encoding less gates in the circuit, which results in far shorter circuits.
Although the \textit{Medium} type introduces more variables compared to the \textit{Dense} type (see \autoref{fig:JointVarsAfterRuntime} top), it features approximately the same circuit depth at bigger problem instances.
Take into consideration that we do not optimise the circuit depth here.
Apart from that, the circuit depth is considerably lower for the \textit{Sparse} type, since it introduces considerably more variables.

\section{Implications on Quantum Software}
\label{sec:effect_on_qc}
In modern software architectures, abstraction layers strive to conceal details about lower layers to allow software engineers to concentrate their work on the relevant architectural state.
In essence, the \textit{SWE/Modelling Domain} and the \textit{Tool/Compiler Domain} are coupled through non-functional requirements, originating from both domains.
It is important to analyse the effect of decisions in one domain to other abstraction layers.
One could argue that more abstraction layers benefit software engineering.
However, as Schönberger \etal~\cite{Schoenberger_2022} showed, this is not necessarily the case for quantum software.
Furthermore, abstraction layers need to be specified without hiding relevant information.
For instance, the runtime limitation of (quantum) algorithms is a cross-abstraction-layer non-functional requirement, for which we showed a substantial increase, when using the \ac{rmd} strategy.
On the other hand, other non-functional requirements are positively influenced by the \ac{rmd} strategy.
Hence, the best fitting balance is to be found, while considering both upstream and downstream influences of decisions in abstraction layers.

By incorporating hardware specifications in advance, we can simultaneously lift constraints about the formulation of \ac{op}s (\ie, using \ac{pubo}s) and optimise for hardware efficient execution.
For this we compared two strategies, namely \emph{map/decompose} (MD) and \emph{reduce/map/decompose} (RMD) with regard to their impact on important metrics of quantum circuits.

Our findings suggest that the use of polynomial order-reduction methods benefits the performance of quantum algorithms, provided that the hardware incorporates sufficiently many qubits.
Therefore, the \ac{qubo} model, which translates to two-qubit interactions, is a non-critical aspect of modelling \ac{op}s from a software engineering perspective.
Thus, the use of higher-order terms to model complex relations is possible, while simultaneously optimising for non-local properties of quantum circuits, namely the circuit depth and the number of gates.
In contrast to that, the mere decomposition of gates, provides a local hardware specific remodelling.
The circuit depth and the number of gates in a circuit are an indicator of the circuit's performance in the \ac{nisq}-era, where decoherence effects and gate errors invalidate the quantum state.

Using variant \textit{Dense} to deploy higher-order \ac{op}s to fully connected quantum hardware such as trapped ion devices \cite{Haffner_2008} achieves relatively low qubit overhead.
For hardware with lower coupling densities, such as
the heavy-hex topology~\cite{Mckay_2023}, the \textit{Sparse} or \textit{Medium} type provide lower densities in the reduced polynomial, while using more qubits. We also see that the \textit{Sparse} variant
focuses its edges among the former variables, which benefits in-homogeneous hardware topologies with concentrated density regions, which also appear in quantum annealing devices~\cite{Hauke2020}.
The mapping and routing of logical quantum circuits to concrete hardware is an \NP{}-complete problem and is thus usually approached with approximation techniques \cite{Siraichi_2018, Yamanaka_2015, Hirata_2009, Cowtan_2014, Zhang_2021}.
Selecting a particular reduction variant to suit the hardware topology can therefore aid in the approximation.
However, further research is needed to assess this influence quantitatively.

\section{Conclusion and Outlook}
\label{sec:concl}
We have discussed that the high-level domain of software engineering and
modelling activities is coupled, through non-functional requirements,
with the low-level domains of compilers, tool-chains and quantum
hardware. This complicates finding appropriate abstraction layers.
By giving an in-depth analysis of a family of automated transformations
from a relatively high-level description of optimisation problems
in PUBO form to circuits for specific quantum hardware, we have
shown that the process allows for exercising control over various
non-functional properties of the resulting computation.

As determined by a numerical experimental analysis, the \ac{rmd} strategy
provides benefits in terms of quantum circuit depth, size and structure, as long as the underlying hardware offers a sufficient amount of qubits
for the problem at hand. As our analysis only depends on the \ac{pbf}'s
structure, the results can be extrapolated to similarly structured \ac{pbf}s.

We provide points of reference that the \textit{Dense} variable-selection type for the \ac{rmd} strategy offers polynomial densities for its degree-2 terms converging to~1, whereas they converge to~0 for the \textit{Sparse} type.
By mapping these different quadratic polynomials (\ie, \ac{qubo}s), resulting from the problem's \ac{pubo}, to a \ac{qaoa} circuit, they influence its depth, size and structure decisively.
As an outlook, we are not limited to the these variable selection types, since the reduction process is of iterative nature.
As both types can be arbitrarily mixed during a reduction process, this allows us to achieve essentially arbitrary degree-2 densities in the resulting polynomial and thus a variety of opportunities to aid in the approximating of the otherwise \NP{}-complete mapping problem~\cite{Siraichi_2018}.

Using reduction methods in an online algorithm demands a higher level of algorithmic optimisation, especially for the \textit{Dense} variable-selection type. This is to ensure that possible quantum advantage of the resulting circuit is not compensated by excessive preparatory classical
effort (nevertheless, it is worth pointing out that the runtime scales in polynomial time for each selection type).
Through the introduction and mathematical analysis of a graph structure, all selection types suggest potential for parallel execution of the transformation, which is a result of the local influence of a reduction step.

\newcommand{\LS}{\censor{LS}\xspace}
\newcommand{\WM}{\censor{WM}\xspace}
\newcommand{\KW}{\censor{KW}\xspace}

\newcommand{\programme}{\blackout{German Federal Ministry of
    Education and Research (BMBF), funding program \enquote{Quantum Technologies---from
      Basic Research to Market}}}
\newcommand{\grantoth}{\censor{\#13N15647 and \#13NI6092}}
\newcommand{\grantsag}{\censor{\#13N16093}}
\newcommand{\hta}{\censor{High-Tech Agenda Bavaria}}

\begin{small}
  \noindent\textbf{Acknowledgements}
  We acknowledge support from \programme, grant \grantoth{} (\LS, \WM) and \grantsag{} (\KW). \WM acknowledges support by the \hta.
\end{small}

\bibliographystyle{IEEEtran}
\bibliography{ms.bib}

\end{document}